\documentclass{sig-alternate-10pt}
\usepackage{graphicx}
\usepackage{subfigure}
\usepackage{algorithmic}
\usepackage{amssymb}
\usepackage{url}
\usepackage{algorithm}
\newcommand{\eg}{{\it e.g.}}
\newcommand{\ie}{{\it i.e.}}
\newcommand{\etc}{{\it etc.}}

\usepackage{color}
\usepackage{ifthen}
\newboolean{TECHREPORT}
\setboolean{TECHREPORT}{false}
\usepackage{cite}
\usepackage{amssymb,amsmath}
\usepackage{newclude}
\usepackage{epstopdf}
\usepackage[font={scriptsize}]{caption}
\usepackage{amsmath,enumerate,psfrag,graphicx,float,bm,theorem,url}
\usepackage{subfig}
\usepackage{booktabs}
\newcommand{\mT}{\mathcal{T}}

\usepackage[T1]{fontenc}

\usepackage{balance}

\usepackage{color}
\definecolor{heraldBlue}{rgb}{0.0,0.0,0.8}
\definecolor{heraldRed}{rgb}{0.8,0.0,0.0}
\definecolor{heraldGray}{rgb}{0.4,0.4,0.4}
\definecolor{heraldBlack}{rgb}{0.0,0.0,0.0} 
\definecolor{heraldGreen}{rgb}{0.0,0.4,0.0} 

\usepackage{tikz}
\usetikzlibrary{arrows,snakes,backgrounds}
\usetikzlibrary{calc}
\usetikzlibrary{positioning}

\begin{document}


\title{From advertising profits to bandwidth prices \\ \LARGE{\emph{A quantitative methodology for negotiating premium peering}}}

\author{
\alignauthor
  \begin{minipage}[t]{0.3\linewidth}
\centering    Laszlo Gyarmati, Nikolaos Laoutaris\smallskip \\
    \affaddr{Telefonica Research}\\
    \email{\normalsize laszlo@tid.es, nikos@tid.es}
  \end{minipage}
  \begin{minipage}[t]{0.22\linewidth}
\centering
Kostas Sdrolias\smallskip\\
\affaddr{Athens University of Economics and Business}\\
        \email{\normalsize sdrolias@aueb.gr}
  \end{minipage}
  \begin{minipage}[t]{0.2\linewidth}
\centering
Pablo Rodriguez\smallskip\\
\affaddr{Telefonica Research}\\
        \email{\normalsize pablorr@tid.es}
      \end{minipage}
  \begin{minipage}[t]{0.25\linewidth}
\centering
Costas Courcoubetis\smallskip\\
\affaddr{Singapore University of
Technology and Design}\\
        \email{\normalsize costas@sutd.edu.sg}
  \end{minipage}
}



\maketitle

\setlength{\emergencystretch}{3em}



We have developed a first of its kind methodology for deriving bandwidth prices for premium direct peering between Access ISPs (A-ISPs) and Content and Service Providers (CSPs) that want to deliver content and services in premium quality. Our methodology establishes a direct link between service profitability, \eg, from advertising, user- and subscriber-loyalty, interconnection costs, and finally bandwidth price for peering. Unlike existing work in both the networking and economics literature, our resulting computational model built around Nash bargaining, can be used for deriving quantitative results comparable to actual market prices. We analyze the US market and derive prices for video that compare favorably with existing prices for transit and paid peering. We also observe that the fair prices returned by the model for high-profit/low-volume services such as search, are orders of magnitude higher than current bandwidth prices. This implies that resolving existing (fierce) interconnection tussles may require per service, instead of wholesale, peering between A-ISPs and CSPs. Our model can be used for deriving initial benchmark prices for such negotiations.

\section{Introduction}

There are many ``tussles''~\cite{Clark2005} affecting the Internet,
ranging across regulation, privacy, network interconnection, and
pricing. The economics of peering~\cite{norton_peering}, is among the thorniest, but yet, least understood ones. The term peering refers to the interconnection between networks for the purpose of exchanging traffic directly between them. Classic \emph{unpaid peering} played a crucial role in the evolution of the Internet. It was usually set up between local access ISPs (or regional tier-2 networks) of similar size for the purpose of avoiding charges and longer paths through upstream ``transit'' providers. The rationale behind such interconnection agreements was that since the involved networks exchanged comparable traffic volumes, there was no issue of who should pay whom. 

This has changed recently with the establishment of peering between dissimilar networks, namely Access ISPs (A-ISPs) and Content and Service Providers (CSPs). Whereas classic unpaid peering between A-ISPs was targeting transit cost reduction, the peering between A-ISPs and CSPs is primarily driven by the need to offer premium quality, \ie, low latency and high bandwidth to traffic and revenue generating services such as video, search, online social networks, and gaming. Such a \emph{premium peering} may include a well provisioned interconnection at various Internet Exchange Points (IXPs), CDN hosting of the CSP's content inside the A-ISP, and even packet level prioritization to avoid congestion at the access part of the network and especially on the last mile. Comcast and Netflix recently signed such a premium agreement, but although there are several (and in many cases controversial) reports attempting to estimate the exact bandwidth price derived by that deal \footnote {{http://blog.streamingmedia.com/2014/02/heres-comcast-netflix-deal-structured-numbers.html}}, the involved parties have not published any official report regarding their agreement.


\vspace{2pt}

\noindent \textbf{The problem.} Classic unpaid peering was justified on the basis of traffic symmetry, which no longer exists since CSPs inject into A-ISPs several orders of magnitude more traffic (\eg, video) than they receive from them (effectively just requests)~\cite{Dhamdhere2010,Labovitz:2010:inter-domain}. This has opened the door to fierce conflicts between A-ISPs and CSPs about who should pay whom and at what rate. The recent disagreement between Free and Google is a good example~\cite{Free_Google}. The main battleground for such conflicts is NANOG.\footnote{\url{www.nanog.org}} Therein peering coordinators have been arguing about payments and have tried to relate them to the question of ``\emph{who benefits the most from the premium peering?}''. They have focused mainly on benefits from reduced transit costs~\cite{norton_peering} and have largely left untouched the question of ``\emph{who can monetize better the superior traffic delivery quality?}''. 

The research literature on networks and economics has developed models of the problem that capture both transit cost savings and monetization of QoE but has derived mostly \emph{qualitative} conclusions. Indeed, as it will be elaborated later (Sect.~\ref{sect:related}) such models miss many of the important details of the conflict, as prices are derived 
based on a bilateral basis rather than in a competitive market. Moreover, they are not driven by real data, and hence, cannot derive \emph{quantitative} results, \ie, actual prices for premium peering. Consequently, such models cannot be validated, nor can they be used for deriving actual prices to be used in peering negotiations.



Motivated by the above, we propose a first-of-its-kind framework capable to analyze premium peerings quantitatively. Our model establishes a direct link between service profitability, distribution costs, and ultimately bandwidth price. It suggests that the way forward and away from existing tussles is to perform per service peering. As with other economic models of the Internet, several questions emerge. Can the model capture the essential dynamics of the Internet ecosystem without becoming computationally intractable? Can the basic parameters of the model be extracted from real world publicly available data? Last, can the results derived by the model be validated against reality?

\vspace{2pt}

\noindent \textbf{Our contributions.} Our main contribution is a methodology leading to a computational model with the properties mentioned above that can be used by A-ISPs and CSPs as a benchmark to derive initial prices for their negotiations around paid peering. We have implemented our model and released a web-site that allows for the analysis of arbitrary markets beyond the one examined in Sect.~\ref{sec:evaluation}. Specific contributions can be broken down as follows:



\noindent $\bullet$ \textbf{Modeling:} Unlike prior work on peering that includes transit ISPs~\cite{Richard2010:shapley}, we focus on the bilateral relationship between A-ISPs and CSPs. In the context of this relationship, we model both costs (\eg, for the extra ports needed for direct peering) and new profits due to improved QoE that translates into increased engagement time of existing customers, and positive (incoming) churn of new customers taken from competitors. Such positive, incoming, churn benefits both the A-ISP and the CSP, whereas the increased engagement time can typically be monetized only by CSPs (since A-ISP offer typically flat contracts for wired broadband~\cite{Laoutaris2008:hotnets}). Capturing ISP- and CSP-side churn poses major modeling challenges. One of our main modeling contribution is a churn model across different services and ISPs that is reasonably close to reality while remaining computationally feasible (Sect.~\ref{sec:churn_model}). We allocate the total surplus by solving a Nash bargaining problem which outputs fair side-payments that can, in turn, be translated into per bit prices for premium peering. The Nash bargaining solution effectively compares the profit of players when they act in isolation vs. when they cooperate and allocates the extra profit accordingly. 



\noindent $\bullet$ \textbf{Data driven approach:} We were able to populate most of the parameters of the proposed model with real-world data. 
A methodological contribution of this work is that populating economic models of the Internet with real data is difficult but not infeasible, at least to a certain level of accuracy that may still produce usable quantitative results. For example we were able to introduce in the model real data for cost of peering, ISP and CSP market shares and profit margins, per service engagement time and sensitivity to QoE, \etc\ (Sect.~\ref{sec:parameterization}). Of course, some parameters are more difficult to set than others, \eg, loyalty factors, so we had for some cases to analyse alternative scenarios with respect to such parameters to check the robustness of our results.  


\noindent $\bullet$ \textbf{Practical takeways:} We run our model for the US market and analyze the results. 
Our findings show that:

\noindent {\em -- CSPs have more ways to monetize improved QoE and thus in most cases they are the ones that pay for premium peering.} Indeed, assuming a similar level of customer loyalty on the A-ISP and CSP markets, the CSP always has the extra benefit of being able to monetize further the premium peering by increasing the engagement time of its existing users and then turning that into additional revenue, through advertising or paid subscriptions. On the other hand, in view of flat broadband contracts, A-ISPs currently do not have any means to monetize improved QoE. 

\noindent {\em -- Large ISPs typically receive payments from CSPs for premium peering whereas smaller ones may offer it for free or even pay for it.} The reason is that the only benefit for ISPs relates to churn. An ISP that connects to the content and services of a popular CSP through a premium peering connection can use the superior QoE to attract customers from competing ISPs. This creates new revenues for the ISP but their extent depends on the market size already captured. A large ISP that already holds a sizable part of the market has a smaller pool from which it could attract new customers. 
If the ISP is small, however, then a premium peering allows it to increase its revenues substantially by attracting a large number of new customers from competitors that do not currently offer premium peering. In that case it is to the best-interest of the small ISP to offer premium peering for free or even pay the CSP for it.     

\noindent {\em -- Customer loyalty and the ability to translate improved QoE into CSP-side revenues affect the volume and even the directionality of payments.} The previous observations assumes a balanced level of loyalty between ISPs and CSPs. If any of the two sides has a substantially higher ``lock-in'' on its customers then this can have a profound impact on the price of premium peering, and even change the directionality of payments. Lock-in's on the ISP can be due to brand power or fixed term contracts with disconnection penalties. For CSPs, lock-in can result from the uniqueness of a service or content, the effort already put in learning and customizing it, or from externalities (\eg., in social networks and gaming). Whatever the reason, if users on one side of the market are more loyal than on the opposite side, then they will reduce the revenues from incoming churn and induce the Nash bargaining solution to compensate by transferring payments from the other side of the market. The final outcome depends jointly on the aforementioned imbalance and the ability of the CSP to monetize QoE. 



\noindent {\em -- Timing matters and first movers have the advantage.} We observe that the post-peering profits for an A-ISP are highly sensitive to what other pairs have established before. The reason is that the first premium peering deals have a larger potential to attract new customers from competitors, whereas late deals are more of a defensive move for retaining new customers rather than bringing new ones. The magnitude of the economic impact is substantial (20\% in our case study).

\noindent {\em -- Balancing the interconnection economics will require per service peering.}  Our case study of the US market indicates that a fair monthly payment from Google to Comcast for premium peering across all services would cost somewhere in the vicinity of \$9.5M and this seems to be stable across much of the parameter space. Breaking down this aggregate payment into per service payments and dividing by the generated bandwidth of each service we were able to derive a \$ value per Gbps per month for each service. It turns out that the fair bandwidth price predicted by our model for video is between \$0.5K--\$1K per Gbps per month which is remarkably close to real transit and paid peering prices from the US market. On the other hand, the Gbps per month \$ value predicted for search, is orders of magnitude higher than current bandwidth prices. This effectively means that current prices are fair for high bandwidth but low revenue per bit services like video but far from fair for high revenue services such as search. Trying to price fairly amorphous bandwidth that includes such strikingly different constituents might be one of the main reasons for the inability to reach a consensus between A-ISPs and CSPs about what constitutes a fair price. A practical takeway from our work is that per service peering might help in resolving the tussle.

\section{The synergy of A-ISP and CSPs}
\label{Sec:AccessCP}
In this section we first explain why we focus on peering issues between A-ISPs and CSPs, then move to define what premium peering is, and finally discuss the consequences of failing to establish one.  

\vspace{2pt}
\noindent \textbf{Why we focus on A-ISPs and CSPs?} Both the A-ISPs and CSPs markets have consolidated in the last decade: the 10 largest telecom operators in the world provide access to more than 2B subscribers~\cite{access_consolidation}, whereas the 10 largest CSPs amount to more than 63\% of the time US users spend online~\cite{cp_consolidation}. Thus, it suffices to establish a small set of agreements for \emph{premium peering} directly between leading A-ISPs and CSPs, guaranteeing that most of the content meets most of the eyeballs through highly provisioned, congestion free, low-delay paths. Furthermore, the content and service discussed here can be replicated in CDNs and clouds close to the end users, from where it is both cheaper and faster to deliver. Combining the ``replicability'' of content and services with
consolidation seems to indicate that \emph{the bulk of content will flow
directly between A-ISPs and CSPs} whereas T-ISPs will handle long-haul
communication (email, voip, \etc), long-tail content distribution
(smaller A-ISPs, CSPs), and backup paths (for both small and large
A-ISPs, CSPs). We therefore focus on
the relationship between A-ISPs and CSPs whose synergy is essential
for high quality content distribution. In the latter category we 
collapse all possible actors such as media creators (\eg, YouTube),
distributors (\eg, NetFlix), CDNs (\eg, Akamai), and content-heavy
tier-1 ISPs (\eg, Level-3).
We model customers and advertisers through their contribution, via payments, to the profits of A-ISPs and CSPs, respectively. 


Since T-ISPs can be bypassed, payments for transit are replaced by payments for direct premium peering between A-ISPs and CSPs. 





\vspace{2pt}
\noindent \textbf{Premium peering.}
%
A premium peering includes some or all of the following: 1) a well provisioned direct peering between the A-ISP and the CSP at multiple points (IXPs, colocation providers, \etc), 2) some form of hosting in CDN nodes deep inside the A-ISP and close to the end users (\eg, next to DSLAMs), which on the one hand improves the quality of delivery, and on the other reduces drastically the delivery costs, and 3) link level prioritization of the said traffic from the point of entry until the customer so that it cannot be obstructed by congestion in the last few links (\eg, edge router to DSLAM, or the last mile). 


\vspace{2pt}
\noindent \textbf{What happens if QoE is not available?}
A-ISPs are claiming that without premium peering revenues they do not have a viable business model~\cite{atkerney10} for upgrading their infrastructures to handle the growth of traffic~\cite{cisco_vni_2012}. In that case, the quality of the content will have to be downgraded to avoid poor Quality of Experience (QoE) due to congestion, \eg, video might have to be encoded at a lower rate or smaller format. For delay sensitive applications such as gaming or browsing of complicated social networks sites, interactivity will be hampered. All of the above have negative consequences for both A-ISPs and CSPs:


\vspace{2pt}
\noindent {\it A-ISP churn.} If an A-ISP does not offer high quality access to the content of a popular CSP, whereas its competitors in the same geographic area do, then it risks losing a number of customers. Such customers would primarily be those that draw high value from accessing the said content.

\vspace{2pt}
\noindent {\it CSP churn.} Failing to establish a premium
connection is detrimental to the CSPs as well. The reason is that
advertisers want to place ads at web sites of CSPs that attract many
eyeballs for substantial time. Indeed, there have been
several studies connecting the QoE with the amount
of time that users spend on a
site~\cite{Kohavi2007:listentocustomers}. Having a non-premium, \ie, best-effort
connection to the eyeballs and, for example, experience pauses while
streaming video, will deter some from spending time at
the site or visiting it at all. Hence, customers and thus ad revenue will move to competitor CSPs that have managed to establish premium peerings with A-ISPs.

In the next sections we build our model for pricing bilateral peering contracts. We start first with modeling the dynamics of churn.




\section{Churn model}\label{sec:churn_model}

In this section, we propose a formal model that captures both A-ISP and CSP churn and also considers the fact  of how important a service is for end users. Before presenting the details of the model, we next describe the essence of our churn model.  We assume that customers switch A-ISPs and CSPs in a probabilistic fashion (\ie, flip coins to decide) according to rules we will disclose later on. Customer churn occurs each time the state of the network connectivity is changed. This involves premium peering to be established (or removed) between an A-ISP and a CSP. In such a case, all the services offered by the given CSP to the customers of the A-ISP, change quality from best-effort to premium (or the reverse). We compute the customer transitions\footnote{Throughout the paper, we use the terms transition, switch, and churn interchangeably.} in two steps. In the first phase, customers decide to switch A-ISPs because other A-ISPs offer some service, for which they really care, in premium quality. In the second phase, customers within the same A-ISP switch CSPs for the services they are interested in and which are now offered with higher quality by a different CSP. Once these two transitions take place the ecosystem is in equilibrium and will be perturbed again when the network connectivity changes. We proceed by defining precisely our churn model for customers. 



\subsection{Definitions}
Let $ISP$ denote the set of A-ISPs on the market while let $CSP$ be the set of CSPs. We arbitrarily order the $K$ different services of the service set $S$
as $(1,2,\ldots,K)$ (\ie, service 1 is video, etc.).
As a first approximation, to describe the service preferences of  a customer, we define its `generic'  type $T$ to be the vector of $(T_1,T_2,\ldots,T_K)$, where the $k$th component
$T_k\in CSP$ defines the CSP providing the $k$th service to the customer of the specific type. Let $\mT$ be the set of possible generic types.
If a customer is of type $T$, we know the CSPs she is using for his services.


We now refine our type definition to include the identity of the most valued service for customers of this generic type (\eg, video is the most valued service of the customer of that generic type). The key idea is to refine the type as little as possible, while capturing the most important aspects of customer differentiation within the same type. The main reason for this is complexity. By doing so we reduce the set of possible customer types, while keeping the `first-order' information for motivating customer transitions between A-ISPs and CSPs. Our new complete definition of a customer type becomes $(i,(s,T))$ where $i\in ISP$, $s\in \{1,\ldots,K\}$ denotes the most valuable service in this  sub-type and $T\in \mT$.  
The  \emph{state} of the type $(i,(s,T))$ corresponds to the number $N(i,(s,T))$ of customers of that type at ISP $i$.

Note that there are at most $K |CSP|^K$ customer types per
ISP. If we were to capture the whole ordering of the popularity of the services instead of keeping the most important one, the number of different types would be larger by a factor of $(K-1)!$. For example, for $|ISP|=4$, $|CSP|=4$, $K=5$, we get about $20000$ types over all ISPs instead of $480000$. 

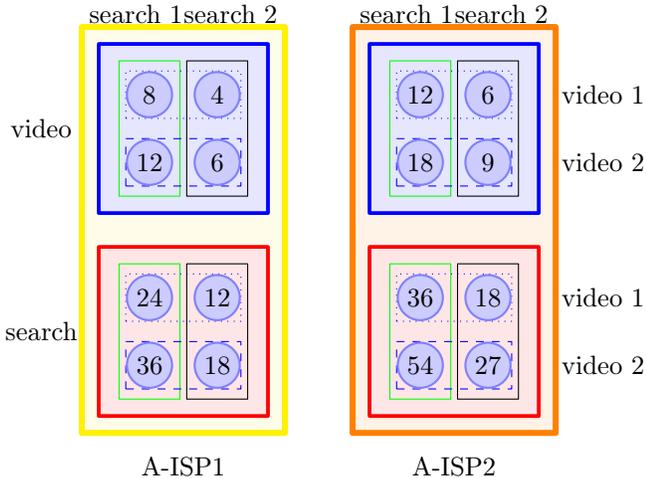
\begin{figure}[tb]
\centering
\tikzstyle{place}=[circle,draw=blue!50,fill=blue!20,thick,inner sep=0pt,minimum size=6mm]
\begin{tikzpicture}[scale=0.9]
	\node[place] (bottomleft) at ( 0,0)  {36};
	\node[place] (bottomright) at ( 1,0)  {18};
	\node[place] (bottommiddleleft) at ( 0,1)  {24};
	\node[place] (bottommiddleright) at ( 1,1)  {12};
	\node[place] (uppermiddleleft) at ( 0,3)  {12};
	\node[place] (uppermiddleright) at ( 1,3)  {6};
	\node[place] (upperleft) at ( 0,4)  {8};
	\node[place] (upperright) at ( 1,4) {4} ; 

	\node[]  at (-0.2,5.2)  {search 1};
	\node[]  at (1.2,5.2)  {search 2};
	\node[]  at (3.8,5.2)  {search 1};
	\node[]  at (5.2,5.2)  {search 2};

	\node[]  at (-1.6,3.5)  {video};
	\node[]  at (-1.6,0.5)  {search};
            
	\node[]  at (6.7,0)  {video 2};
	\node[]  at (6.7,1)  {video 1};
	\node[]  at (6.7,3)  {video 2};
	\node[]  at (6.7,4)  {video 1};
	        
	\node[]  at (0.5,-1.5)  {A-ISP1};
	\node[]  at (4.5,-1.5)  {A-ISP2};

	\node[place] (rbottomleft) at ( 4,0)  {54};
	\node[place] (rbottomright) at ( 5,0)  {27};
	\node[place] (rbottommiddleleft) at ( 4,1)  {36};
	\node[place] (rbottommiddleright) at ( 5,1)  {18};
	\node[place] (ruppermiddleleft) at ( 4,3)  {18};
	\node[place] (ruppermiddleright) at ( 5,3)  {9};
	\node[place] (rupperleft) at ( 4,4)  {12};
	\node[place] (rupperright) at ( 5,4)  {6} ; 

	\begin{pgfonlayer}{background}
		\filldraw [line width=0.8mm,join=round,fill=yellow!10,draw=yellow] (-1,-1) rectangle (2,5);
		\filldraw [line width=0.5mm,join=round,fill=red!10,draw=red] (-0.75,-0.75) rectangle (1.75,1.75);
		\draw [join=round,draw=green] (-0.45,-0.5) rectangle (0.45,1.5);
		\draw [join=round,draw=black] (0.55,-0.5) rectangle (1.45,1.5);
		\draw [join=round,draw=blue,dashed] (-0.35,-0.35) rectangle (1.35,0.35);
		\draw [join=round,draw=blue,dotted] (-0.35,0.65) rectangle (1.35,1.35);
		\filldraw [line width=0.5mm,join=round,fill=blue!10,draw=blue] (-0.75,2.25) rectangle (1.75,4.75);
		\draw [join=round,draw=green] (-0.45,2.5) rectangle (0.45,4.5);
		\draw [join=round,draw=black] (0.55,2.5) rectangle (1.45,4.5);
		\draw [join=round,draw=blue,dashed] (-0.35,2.65) rectangle (1.35,3.35);
		\draw [join=round,draw=blue,dotted] (-0.35,3.65) rectangle (1.35,4.35);

		\filldraw [line width=0.8mm,join=round,fill=orange!10,draw=orange] (3,-1) rectangle (6,5);
		\filldraw [line width=0.5mm,join=round,fill=red!10,draw=red] (3.25,-0.75) rectangle (5.75,1.75);
		\draw [join=round,draw=green] (3.55,-0.5) rectangle (4.45,1.5);
		\draw [join=round,draw=black] (4.55,-0.5) rectangle (5.45,1.5);
		\draw [join=round,draw=blue,dashed] (3.65,-0.35) rectangle (5.35,0.35);
		\draw [join=round,draw=blue,dotted] (3.65,0.65) rectangle (5.35,1.35);
		\filldraw [line width=0.5mm,join=round,fill=blue!10,draw=blue] (3.25,2.25) rectangle (5.75,4.75);
		\draw [join=round,draw=green] (3.55,2.5) rectangle (4.45,4.5);
		\draw [join=round,draw=black] (4.55,2.5) rectangle (5.45,4.5);
		\draw [join=round,draw=blue,dashed] (3.65,2.65) rectangle (5.35,3.35);
		\draw [join=round,draw=blue,dotted] (3.65,3.65) rectangle (5.35,4.35);	  \end{pgfonlayer}
\end{tikzpicture}
\caption{Toy example: the distribution of users among different types based on their most important service, their A-ISPs, and their CSPs used for the different services. The $y$-axis corresponds to the preferred services. For example, there are 27 users of A-ISP2 who are interested in search, and use CSP-2 as their search and video provider (bottom left type). Similarly, 4 users have A-ISP1 as access provider, their most important service is video, they use CSP2 as their search provider, and they use the video services of CSP1 (second type in the top left).}
\label{fig:ptype_toy_example}
\end{figure}

\vspace{4pt}
\noindent \textbf{Toy example. }
We illustrate our ecosystem and the types of the customers in a toy example. Let us consider a market where there are two A-ISPs, two CSPs, and two services, namely, search and video. Let us assume that this market has 300 customers in total. We assume that A-ISP1 has 2/5 of the access market, CSP1 provides search for 2/3 of the users and its video service is used by 2/5 of the users. We further assume that search is the most important service for 3/4 of the users. Distributing the customers based on these ratios yields the state of the different types as we depict in Figure~\ref{fig:ptype_toy_example}. The figure also reveals---via its color-coding---the way how the users can be summarized for a specific service, for a given service of a CSP, or for a CSP itself.

\subsection{Transition of customers}
We define now the transitions between the states of  customer types. Such transitions occur (i) because customers find their preferred service being offered by the same CSP at premium quality at a new A-ISP, or (ii) because at their current A-ISP the quality of their preferred service became available at a higher level by an other CSP. Hence, the transitions of the customers are motivated by \emph{changes of quality only regarding the customers' preferred service}. In all cases, a customer's preferred service category remains the same (\eg, if video is the customer's most valued service, this remains the case during all the transitions), whereas the customer may change its A-ISP or CSP. Therefore, in case of (i), a customer is keeping the same $(s,T)$ and just switches its A-ISPs. In case (ii), the customer changes type within the same A-ISP, by only changing the CSP of her preferred service. By defining the corresponding transition probabilities, the transitions of the various types are straightforward since we keep track the number of customers of each type per A-ISP. We next describe the customer churn  if a premium peering is established.
 \medskip




\noindent \textbf{Phase 1: churn across A-ISPs. }
If A-ISP $j$ and CSP $x$ establish a premium peering, the customers transition from type $(i,(s,T))$ to  type $(j,(s,T))$. Let $n_1 = N(i,(s,T))$ and $n_2 = N(j,(s,T))$ denote the number of customers that A-ISP $i$ and $j$ have in the given types, respectively. We determine the transition of the customers along the following rules:
\begin{enumerate}
\item The transition of customers occurs only if $T_s = x$ and A-ISP $i$, had no premium peering with CSP $x$. 
\item The probability of transition is $\gamma=(1-\beta(i)) h(s)$, where $\beta(i)$ is the stickiness (\ie, loyalty) to A-ISP $i$ and $h(s)$ is the probability of customers who mainly care about service $s$ (\eg, video) to switch ISPs because the quality of $s$ is improved. 
\item As a result of the state transition, the number of churning customers is $\Delta=n_1 \gamma$, hence the new states become $N(i,(s,T)) = n_1 - \Delta$ and\\
 $N(j,(s,T),) = n_2 + \Delta$, respectively.
\end{enumerate}

\noindent \textbf{Phase 2: churn across CSPs. }
If A-ISP $i$ and CSP $x$ establish a premium connection, the customers churn from type $(i,(s,T))$ to type $(i,(s,T'))$, where $T'_s=x$ and $T_s\neq y$ hold. If the starting states of these types are $N(i,(s,T)) = n_1$ and $N(i,(s,T')) = n_2$, we compute the volume of churning customers as follows: 
\begin{enumerate}
\item The transition of customers occurs only if CSP $T_s$ has no existing premium peering agreement with A-ISP $i$. 
\item The probability of the churn is $\gamma=(1-\theta(T_s)) g(s)$ where $\theta(T_s)$ is the stickiness to CSP $T_s$ and $g(s)$ is the probability for customers that mainly care about service $s$ (e.g., video) to switch CSPs because the quality of $s$ improves. 
\item The number of transitioning customers is $\Delta=n_1 \gamma$; the new states 
become $N(i,(s,T)) = n_1 - \Delta $ and $N(i,(s,T')) = n_2 + \Delta$, respectively.
\end{enumerate}

\noindent \textbf{Toy example. }
We use the same market we depicted in Figure~\ref{fig:ptype_toy_example} to illustrate the steps of our churn model. The value of the states represent the number of customers after the churn, while the labels of the arrows indicate the volume of switching customers. We assume that 1/3 of the customers switch their A-ISPs while a CSP with premium services captures 1/2 of the users. If A-ISP1 and CSP1 set up a premium peering, due to premium quality customers of A-ISP2 churn to A-ISP1 (phase 1). In the figure we show which types are affected: video-enthusiastic customers who are using the video service of CSP1, and search-eager customers who are using the search engine of CSP1. The second effect of the premium peering is that some customers of A-ISP1 change their current CSP to access their most important service in premium quality. We show with dashed arrows the types from which customers churn: a fraction of the customers who care for video (search) churn from CSP2 to the video (search) service of CSP1.


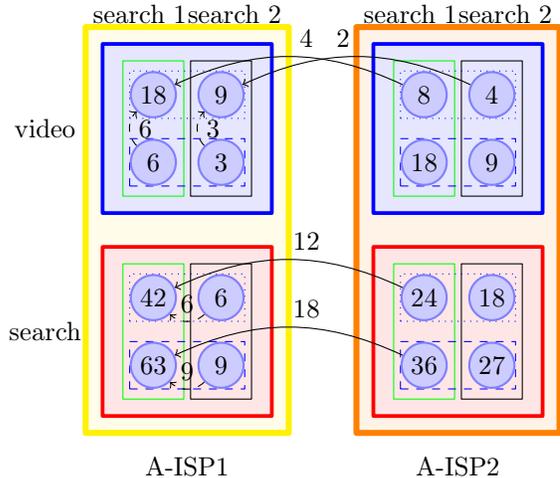
\begin{figure}[tb]
\centering
\tikzstyle{place}=[circle,draw=blue!50,fill=blue!20,thick,inner sep=0pt,minimum size=6mm]
\begin{tikzpicture}[scale=0.9]
	\node[place] (bottomleft) at ( 0,0)  {63};
	\node[place] (bottomright) at ( 1,0)  {9}  edge [->,bend left=45,dashed] node[auto,swap] {9} (bottomleft);
	\node[place] (bottommiddleleft) at ( 0,1)  {42};
	\node[place] (bottommiddleright) at ( 1,1)  {6}   edge [->,bend left=45,dashed] node[auto,swap] {6} (bottommiddleleft);
	\node[place] (uppermiddleleft) at ( 0,3)  {6}  edge [->,bend left=45,dashed] node[auto,swap] {6} (upperleft);
	\node[place] (uppermiddleright) at ( 1,3)  {3}  edge [->,bend left=45,dashed] node[auto,swap] {3} (upperright);
	\node[place] (upperleft) at ( 0,4)  {18};
	\node[place] (upperright) at ( 1,4) {9} ; 

	\node[]  at (-0.2,5.2)  {search 1};
	\node[]  at (1.2,5.2)  {search 2};
	\node[]  at (3.8,5.2)  {search 1};
	\node[]  at (5.2,5.2)  {search 2};

	\node[]  at (-1.6,3.5)  {video};
	\node[]  at (-1.6,0.5)  {search};
            
	        
	\node[]  at (0.5,-1.5)  {A-ISP1};
	\node[]  at (4.5,-1.5)  {A-ISP2};

	\node[place] (rbottomleft) at ( 4,0)  {36} edge [->,bend right=25] node[auto,swap] {~~~~18} (bottomleft);
	\node[place] (rbottomright) at ( 5,0)  {27};
	\node[place] (rbottommiddleleft) at ( 4,1)  {24} edge [->,bend right=25] node[auto,swap] {~~~~12} (bottommiddleleft);
	\node[place] (rbottommiddleright) at ( 5,1)  {18};
	\node[place] (ruppermiddleleft) at ( 4,3)  {18};
	\node[place] (ruppermiddleright) at ( 5,3)  {9};
	\node[place] (rupperleft) at ( 4,4)  {8} edge [->,bend right=25] node[auto,swap] {~~~~4} (upperleft);
	\node[place] (rupperright) at ( 5,4)  {4} edge [->,bend right=25] node[auto,swap] {2~~~~} (upperright);

	\begin{pgfonlayer}{background}
		\filldraw [line width=0.8mm,join=round,fill=yellow!10,draw=yellow] (-1,-1) rectangle (2,5);
		\filldraw [line width=0.5mm,join=round,fill=red!10,draw=red] (-0.75,-0.75) rectangle (1.75,1.75);
		\draw [join=round,draw=green] (-0.45,-0.5) rectangle (0.45,1.5);
		\draw [join=round,draw=black] (0.55,-0.5) rectangle (1.45,1.5);
		\draw [join=round,draw=blue,dashed] (-0.35,-0.35) rectangle (1.35,0.35);
		\draw [join=round,draw=blue,dotted] (-0.35,0.65) rectangle (1.35,1.35);
		\filldraw [line width=0.5mm,join=round,fill=blue!10,draw=blue] (-0.75,2.25) rectangle (1.75,4.75);
		\draw [join=round,draw=green] (-0.45,2.5) rectangle (0.45,4.5);
		\draw [join=round,draw=black] (0.55,2.5) rectangle (1.45,4.5);
		\draw [join=round,draw=blue,dashed] (-0.35,2.65) rectangle (1.35,3.35);
		\draw [join=round,draw=blue,dotted] (-0.35,3.65) rectangle (1.35,4.35);

		\filldraw [line width=0.8mm,join=round,fill=orange!10,draw=orange] (3,-1) rectangle (6,5);
		\filldraw [line width=0.5mm,join=round,fill=red!10,draw=red] (3.25,-0.75) rectangle (5.75,1.75);
		\draw [join=round,draw=green] (3.55,-0.5) rectangle (4.45,1.5);
		\draw [join=round,draw=black] (4.55,-0.5) rectangle (5.45,1.5);
		\draw [join=round,draw=blue,dashed] (3.65,-0.35) rectangle (5.35,0.35);
		\draw [join=round,draw=blue,dotted] (3.65,0.65) rectangle (5.35,1.35);
		\filldraw [line width=0.5mm,join=round,fill=blue!10,draw=blue] (3.25,2.25) rectangle (5.75,4.75);
		\draw [join=round,draw=green] (3.55,2.5) rectangle (4.45,4.5);
		\draw [join=round,draw=black] (4.55,2.5) rectangle (5.45,4.5);
		\draw [join=round,draw=blue,dashed] (3.65,2.65) rectangle (5.35,3.35);
		\draw [join=round,draw=blue,dotted] (3.65,3.65) rectangle (5.35,4.35);	  \end{pgfonlayer}
\end{tikzpicture}
\vspace{-12pt}
\caption{The churn of customers between types if A-ISP1 and CSP1 establish a premium peering. Phase 1: churn from A-ISP2 to A-ISP1. Phase 2 (dashed arrows): churn from CSP2 to CSP1.}
\label{fig:churn_toy_example}
\end{figure}

\section{Computing the price of  premium peering}\label{sec:Nash_bargaining}

\begin{figure}[tb]
\centering
\begin{tikzpicture}
\node[circle,draw] {T-ISP}
[sibling distance=6cm]
child {node[circle,draw] (left node) {A-ISP i}
	[sibling distance=2cm]
			edge from parent [dashed,line width=2pt]
			node[left] {$t(i)$~~}
}
child {node[circle,draw] (right node) {CSP x}
	[black,sibling distance=2cm]
		edge from parent [dashed,line width=2pt]
		node[right] {~~$t(x)$}
};
\coordinate [label=above:$c_p(i)$] (lp1) at (-1.5,-2.1);
\coordinate [label=above:$c_p(i)$] (lp2) at (1.7,-2.1);
\coordinate [label=above:$u(i)\textrm{,} n(i) $] (l3) at (-3,-2.75);
\coordinate [label=above:$R(s) \textrm{,} a(s) \textrm{,} \rho(s) \textrm{,} \tau(s) \textrm{,} \varphi(s) \textrm{,}$] (l3) at (2,-2.75);
\draw[black, line width=2pt,rounded corners] (left node) -- (right node);
\end{tikzpicture}
\caption{A premium peering scenario with the parameters of the actors. The traffic flows directly between A-ISP $i$ and CSP $x$ if they set up premium peering.}
\label{fig:bilateral_market}
\end{figure}
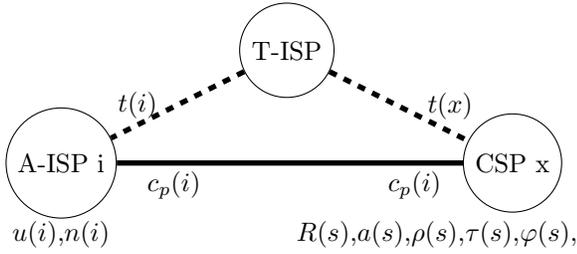

We now introduce an economic framework for computing fair peering prices. Our analysis captures the revenues and the expenditures of an ISP and a CSP both before and after they establish a premium peering agreement, taking also into account customer churn.
We illustrate the market scenario along with the appropriate parameters of the stakeholders in Figure~\ref{fig:bilateral_market}. Initially, the traffic flows between A-ISP $i$ and the CSP $x$ throughout a transit ISP (T-ISP), while under a premium peering this traffic will flow directly between them. To analyse the economic gains from premium peering we need to define the revenues and expenditures of the actors before and after the agreement.
We use the `$\hat{~}$' notation to distinguish the various quantities after peering from the respective values before peering is established . 

\vspace{15pt}
\subsection{Profits before peering}

For A-ISP $i$, let $u(i)$ denote the profit per customer, $n(i) = \sum_{s,T}N(i,(s,T))$ the total number customers, and  
$n(i)_{\xi=x} = \sum_{s,T | T_{\xi} = x} N(i,(s,T))$ the number of customers that subscribe for service $\xi$ to CSP $x$.

CSPs have two main sources of income: subscription fees and advertisement-based revenue. If $R(s)$ = profit per customer subscribing to service $s$ (assumed uniform across CSPs), the total subscription related profit of CSP $x$ 
from the customer base of ISP $i$ is $\sum_{\xi \in S} R(\xi) n(i)_{\xi=x}$.

The advertisement profit of a CSP is a product of $\tau(s)$, the engagement time (\eg, minutes) of the customers, \ie, the time the customers spend using service $s$, $\rho(s)$, the rate of the clicks\footnote{or impressions depending on the scheme of the advertisement campaign} of the advertisements in case of service $s$, and $a(s)$, the profit per click in case of service $s$.

Without a peering agreement, A-ISP $i$ and CSP $x$ have to pay for transit, being charged the product of the traffic volume and the unit price of the transit service. Let $\varphi(s)$, $t(i)$, $t(x)$ denote the average traffic rate required per user engagement time of service $s$ and the unit cost of transit (\eg, per Mbps) for A-ISP $i$ and CSP $x$, respectively.
%
Then, summing up all the traffic originating from the services of CSP $x$ and terminating at A-ISP $i$, the total expenditure of A-ISP $i$ is:
\begin{equation}
	c_t (i)= t(i) \sum_{\xi \in S}  \varphi(\xi) \tau(\xi) n(i)_{\xi=x} \nonumber \, ,
\end{equation}
and similarly for CSP $x$ the transit cost is
\begin{equation}
	c_t(x)= t(x) \sum_{\xi \in S}  \varphi(\xi) \tau(\xi) n(i)_{\xi=x} \nonumber \, .
\end{equation}

Based on the above, the profits of A-ISP $i$ and CSP $x$ before the peering agreement are respectively
$V_i = n(i) u(i)- c_t(i)$ and $V_x = \sum_{\xi \in S} \left(R\left(\xi\right) + a\left(\xi\right) \rho\left(\xi\right) \tau\left(\xi\right)\right) n(i)_{\xi=x} - c_t(x)$.
These expressions contain solely the monetary aspect of the relation between the two actors $i,x$, \ie, they do not include the revenues and expenditures related to the other A-ISPs and CSPs present in the market since these are not affected by premium peering.

\subsection{Profits after  peering}

If the A-ISP and the CSP agree to engage in a premium peering, their profits change due to three reasons. First, the cost of transmitting the data between the parties changes. They do not incur  anymore transit cost  but need to pay for the new costs of peering $c_p(i)$ and $c_p(x)$ respectively, for A-ISP $i$ and CSP $x$ \footnote{The peering costs of the actors are not necessarily identical although both parties handle the same amount of traffic. For example, one of them may face a collocation cost to build into the peering facility (\eg, into a public IXP) where the other party is already present. The cost of the peering depends on several factors like the volume of the traffic, collocation needs, and whether the peering agreement includes a CDN service provided by the A-ISP. }.

Second, customers are becoming more engaged with the services of the CSP if the services are accessible in premium quality. The enhanced quality may affect all the parameters of the CSPs' revenue streams: the profit per customer paying subscription fees ($\hat{R}(s)$), the engagement time of the customers ($\hat{\tau}(s)$), the rate of clicks ($\hat{\rho}(s)$), and the profit per click ($\hat{a}(s)$). Moreover, enhanced quality services may impose additional traffic loads ($\hat{\varphi}(s)$). On the A-ISP side, the profit per customer could also change due to the improved quality ($\hat{u}(i)$).

Third, the improved quality acts as a driving force for the churn of the customers, as we discussed it in details in Section~\ref{sec:churn_model} for calculating the new values of $\hat{N}(i,(s,T))$. Our post-peering values of the profits for A-ISP $i$ and CSP $x$ become
$\hat{V}_i = \hat{n}(i) \hat{u}(i) - c_p(i)$ and 
$\hat{V}_x = \sum_{\xi \in S} \left(\hat{R}\left(\xi\right) + \hat{a}\left(\xi\right) \hat{\rho}\left(\xi\right) \hat{\tau}\left(\xi\right) \right) \hat{n}(i)_{\xi=x}- c_p(x)$ respectively.
 
\subsection{Nash bargaining solution}
Based on the profits of the actors before and after the premium peering, 
we quantify the total monetary benefit of the peering agreement as
\begin{equation}
	U = \hat{V}_i - V_i + \hat{V}_x - V_x \, .
\end{equation}
If $U$ is negative, then trivially there is no sense to make such an agreement, and if positive, as result of their cooperation,
it should be shared among them.
We use the Nash Bargaining solution~\cite{fudenberg-tirole} as a concept to define fair profit allocations when the possible actions
of the players are to establish the peering connection at a specific level of QoS or not peer at all. We assume that each player has an equal urge to
settle the negotiations and that arbitrary monetary transfers
are possible (i.e., that after peering the players can pay each other any amount that they have agreed upon). This assumption to allow for an arbitrary payment among the actors simplifies our solution concept tremendously; the solution is to peer if there are positive total profits, and compute the monetary transfers to ensure that each player obtains equal extra profits computed on top of her fall back solution profits (not peer at all, if negotiations brake). Formally, we like to find the fair new \emph{total} profits $z_i$ and $z_x$ for the two actors by solving
\begin{equation}
	\max_{z_i+z_x = U}\left(z_i - V_i\right)\left(z_x - V_x\right)\, .
\end{equation}
Once we obtain the net fair profits, we compute the respective payment from CSP to ISP as $\hat{V}_x - z_x$, which can be positive or negative.
The interpretation is that if under a fair arbitration, CSP $x$ should end up with a net profit  of $z_x$ that is less than her realised profit 
 $\hat{V}_x$ after peering (with no side payments to the ISP), he should turn back to the ISP the part of the profit that is above the fair amount.
The same results values but with an opposite sign follow if we consider the case of the payment from the ISP to the CSP. 
In our case the fair payment $w_x$ is 
\begin{equation}
	w_x = \frac{1}{2} \left[ (\hat{V}_x -V_x) -(\hat{V}_i -V_i) \right] \, .
\end{equation}

 The solution definition guarantees 
 that if the peering results to a positive surplus $U \geq 0$, none of participants end up with less profit than they would realise before the establishment of the premium peering.

\section{Parameterization of the model} \label{sec:parameterization}
In this section, we apply the proposed model quantitatively to investigate the derived premium peering prices in the U.S. Internet ecosystem. 

Based on real datasets, we present the assumptions we make on the parameters of the model. Initially, we distinguish two types of assumptions: modeling and parameterization assumptions. The modeling assumptions reduce the the parameter state size by claiming sameness of some metrics, while for the parameterization assumptions we use real-world datasets in order to assign reasonable values to the model's parameters. Furthermore, we classify the various assumptions in four different sets: Services assumptions (Section 5.1), CSPs assumptions (Section 5.2), A-ISPs assumptions (Section 5.3)and General market assumptions (Section 5.4). In the following Tables (Table~\ref{tab:services_assum}, Table~\ref{tab:csps_assum}, Table~\ref{tab:aisps_assum} and Table~\ref{tab:gen_assum}), we summarize the considered assumptions, before proceeding to the justification of our parameterization. 

\begin{table}[!htbp]
\small
\centering
\begin{tabular}{|c||c|}
\hline
 Services' parameters& subsection\\
\hline \hline
Probability of A-ISP churn& 5.1.1	\\	\hline
Probability of CSP churn& 5.1.1\\	\hline	
Before peering traffic rates& 5.1.2\\	\hline
After peering traffic rates& 5.1.2\\	\hline
Before peering engagement times& 5.1.2\\	\hline
After peering engagement times& 5.1.2\\	\hline
Relative Importance& 5.1.2\\	\hline
\end{tabular}

\caption{Modeling and parameterization assumptions for services.}\label{tab:services_assum}
\end{table}

\begin{table}[!htbp]
\small
\centering
\begin{tabular}{|c||c|}
\hline
 CSPs' parameters& subsection\\
\hline \hline
CSP stickiness assumptions& 5.2.1	\\	\hline
Post peering advertising rates& 5.2.1\\	\hline	
Market shares& 5.2.2\\	\hline
Estimation of CSP stickiness& 5.2.2\\	\hline
\end{tabular}

\caption{Modeling and parameterization assumptions for CSPs.}\label{tab:csps_assum}
\end{table}

\begin{table}[!htbp]
\small
\centering
\begin{tabular}{|c||c|}
\hline
 A-ISPs' parameters & Subsection\\
\hline \hline
A-ISP stickiness assumptions& 5.3.1	\\	\hline
Number of subscribers and annual profits& 5.3.2\\	\hline	
Estimation of CSP stickiness& 5.3.2\\	\hline
\end{tabular}

\caption{Modeling and parameterization assumptions for A-ISPs.}\label{tab:aisps_assum}
\end{table}

\begin{table}[!htbp]
\small
\centering
\begin{tabular}{|c||c|}
\hline
General parameters & Subsection\\
\hline \hline
Market conditions& 5.4.1	\\	\hline
Transit and peering costs& 5.4.2\\	\hline	
\end{tabular}

\caption{General parameters}\label{tab:gen_assum}
\end{table}

\subsection{Services}

We assume that a premium peering deal includes at most, the following four types of services: video, online social network, search, and gaming. This assumption is motivated by the fact that the aforementioned services cover most of the time an average end-user spends online, while capturing more than 70\% of the online advertisement-based revenues in 2013 ~\cite{iab_global_2014}, under the assumption that Display/Banner and Rich Media ad-formats, fall in any of these types of services. 

Furthermore, we distinguish the video services in two sub-types: user-centric video, and commercial based video services such as online movies and TV shows. The main reason for this distinction is that user-centric video is mainly delivered by CSPs whose profits are based on advertisements, while CSPs that deliver movies or TV series are financially based on monthly subscription fees. Moreover the duration (and hence the end-users' engagement times) for user-centric content is significantly smaller in comparison with online movies.

\subsubsection {Modelling assumptions}
\vspace{4pt}
Initially we assume that end-users' favorite type of service falls in any of the aformentioned service categories, \eg, can be either search or video or osn or gaming. Furthermore,  we attempt to capture  the inherent characteristics of each service, that differentiates itself from all the other services. Thus, we assume that the probabilities of churn are not equal for the various services, but increasing in a linear manner, such that the sum of the churn probabilities of the various services is equal to 1.

\noindent \textbf{Probability of A-ISP churn.} The probability of the A-ISP churn  ($h(s)$), increases in the following order of the services: search, video, OSN, gaming. In case of gaming, two factors promote the churn of a user to a new A-ISP: the network effect of its friends playing with and the cost of the gaming console. In case of OSN, the network effect increases the churn probability, while the power of a video service is related to the variety of its available content. Finally, the search engines can be replaced in the easiest way as far as the usability aspect is concerned, as long as the search results are relevant.
As explained before, the increase is linear: $h(search) + 3\mu = h(video) + 2\mu = h(osn) + \mu = h(gaming) = h$. We use $h=0.4$ and $\mu=0.1$ as parameters.\vspace{4pt}

\noindent \textbf{Probability of CSP churn.} Similar trends are considered in case of the probability of the CSP churn ($g(s)$). However, in this case the order of the services reverses, \eg, replacing the search engine has the highest probability. Here we assume again a linear relation between the services: $g(gaming) + 3\nu = g(osn) + 2\nu = g(video) + \nu = g(search) = g$. We use $g=0.4$ and $\nu=0.1$ as parameters.

\subsubsection {Parameterization assumptions} \label{sub:par_assumptions}
We use real-world datasets in order to assign reasonable values to the model's parameters. Hence, we apply the forthcoming parameterization to analyze the peering decisions of A-
ISPs and CSPs. 

\noindent \textbf{Importance of services.} As Nielsen ~\cite{nielsenwire2010} points out, in 2010 end-users spent 6.01 minutes for online gaming services per day. Assuming that online gaming hours up 6\% year-over-year ~\cite{vg247}, it appears that current online gaming engagement times are 7.45 minutes per day.

We use the minutes end-users spend on Netflix, Youtube, Google search and Facebook as proxies for estimating the daily engagement times for commercial-based video, user-centric video, search and online social network services respectively.
Based on ~\cite{stat_brain}, it turns out that on average, end-users spend 11.8 minutes per day on Youtube and 15.2 minutes per day on Facebook. In the case of online movies, it has been  estimated that subscribers of Netflix spend around 103 minutes per day on online views ~\cite{netflix_time}. As the market coverage of Netflix in the U.S. market is around 38\% ~\cite{netflix_share}, we conclude that on average, users spend 39.14 minutes per day on online movies. Finally for online search, we assume that end-users spend around 6.72 minutes of their online time per day (see for example ~\cite{infographic}).
	
Given that the time the customers spend online describes accurately the importance of the service categories, we present in Table~\ref{tab:services_time}, the relative importance of each service.

\begin{table}[tb]
	\small
\centering
\begin{tabular}{|c||c|}
\hline
Service & Relative importance \\
\hline
\hline 
User-centric Video &  14.69\% \\
\hline
Online social networks & 18.95\% \\
\hline
Search & 8.36\% \\ 
\hline
Gaming & 9.27\% \\ 
\hline
Commercial-based video & 48.73\% \\
\hline
\end{tabular}
\caption{Importance of services based on spent time} \label{tab:services_time}
\end{table}



\noindent \textbf{Traffic rates before a premium peering agreement.} As major online video providers (such as Netflix and Amazon), use Adaptive Bit Rate streaming \footnote{\url{http://scenic.princeton.edu/network20q/wiki/index.php?title=Adaptive_Bitrate_Streaming}}, we assume that the average download traffic rates for commercial-based video, are on average 22.5 MB/min (average DVD quality). Based on \footnote{\url{http://sonnati.wordpress.com/2011/09/15/bandwidth-is-running-out-lets-save-the-bandwidth}} and assuming that on average end-users watch user-centric video services on 480p encoding, or 7.5 MB per minute of engagement time.

For computing the traffic rates of online social network services we user the report of Sandvine \cite{sand2014}. The median monthly consumption of north american fixed accesses was 19.4 GB, and during the peak hours Facebook accounted for the 1.99\% of the total traffic, and hence we assume that the average traffic for this type of service is 0.84 MB/min per user.

The traffic of online gaming is 0.12\% of the traffic of video according to \cite{cisco_vni_2012}, and assuming similar traffic volume for search as the gaming one, it is derived that the traffic rates for online gaming are 0.051 MB/min, and 0.054 MB/min for search.



	%

\noindent \textbf{Post-peering traffic rates.} We assume that the post peering quality of the user-centric video increases from 480p to 720p, resulting  2.1-time larger traffic rate: $\hat{\varphi}(user-centric \hspace{1.3 mm} video) = 2.1 \varphi(user-centric \hspace{1.3 mm} video)$. Furthermore, as end-users are more sensitive on the offered QoE of the online movies, we consider 4 times increase of the post peering traffic rates: $\hat{\varphi}(commercial-based \hspace{1.3 mm} video) = 4 \varphi(commercial-based \hspace{1.3 mm} video)$. All the other traffic rates remain the same, as for these services, the premium quality means dominantly reduced latency: $\hat{\varphi}(s) = \varphi(s), \, \forall s \in \{\textrm{OSN,search,gaming}\}$.

\subsection{CSPs}
We consider two types of CSPs: Ad-powered CSPs, which do not offer subscription-fee based services, \ie, their revenues are generated solely via advertisements: $R(s)=0, \hat{R}(s)=0,  \, \forall s \in S$, and subscription-based CSPs which charge end-users a fixed fee per month., although they may also earn additional advertisement-based profits.

\subsubsection {Modeling assumptions}

\noindent \textbf{CSP stickiness. }
We assume that the stickiness of the CSPs is identical: $\theta(x)=\theta, \, \forall x \in CSP$.

\noindent \textbf{Post peering advertising rates. }For any type of CSP, we assume that premium quality does not impact the click through rates and the per click revenues of the services: $\rho(s)=\hat{\rho}(s), a(s)=\hat{a}(s), \, \forall s \in S$.

Finally, a CSP $x$ is able to capture churning customers only if it had already customers for the given $\xi$ service: $\sum_{i,s,T_{\xi} = x} N(i,(s,T)) > 0$.

\subsubsection {Parameterization assumptions}

\noindent \textbf{Market shares for ad-powered CSPs.}  We rely on the reports of comScore on the popularity of search and video services ~\cite{comscore_search2014} and ~\cite{comscore_video2014}, while for osn services we use a report published by socialfresh.com ~\cite{socialfresh2014}. More specifically for Google, we sum the popularity of Google+ and Blogger because Blogger is owned by Google. Owed to the lack of specific reports, we assume a uniform distribution of customers in case of games taking into account the fact that all the CSPs have massive gaming services. We distribute the end users among the CSPs for each service category, proportionally to the time they spend using the given service, and summarize our findings in Table~\ref{tab:csp_user_distributionk}.
\begin{table}[tb] 
	\small
\centering
\begin{tabular}{|c||c|c|c |c|}
	\hline
	CSP & video \cite{comscore_video2014} & OSN \cite{socialfresh2014} & search \cite{comscore_search2014} & gaming* \\
	\hline
	\hline
	Google & 39.39\% & 29.6\% & 69.16\% & 20\% \\
	\hline
	Microsoft & 8.89\% & 0\% & 18.85\% & 20\% \\
	\hline
	Yahoo & 13.19\% & 4.06\% & 10.56\% & 20\% \\
	\hline
	Facebook & 22.01\% & 68.95\% & 0\% & 20\% \\
	\hline
	AOL & 16.52\% & 0\% & 1.43\% & 20\% \\	
	\hline
\end{tabular}
\caption{Customer distribution among the services of the ad-powered CSPs (*assumption). Video briefly refers to user-centric video services.}
\label{tab:csp_user_distributionk}
\end{table}

\noindent \textbf{Market shares for subscription-based CSPs.} We use a report fron Nielsen published in 2013, to distribute U.S end-users across the considered subscription-based CSPs~\cite{netflix_share}. Netflix holds 38\% of the market, Amazon Prime 13\% and Hulu Plus 6\% (see Table ~\ref{tab:csp_user_distributionl}).
\begin{table}[tb]
	\small
\centering
\begin{tabular}{|c||c|}
	\hline
	CSP & online movies \cite{netflix_share} * \\
	\hline
	\hline
	Netflix & 38\% \\
	\hline
	Amazon Prime & 13\% \\
	\hline
	Hulu Plus & 6\% \\
	\hline
\end{tabular}
\caption{Customer distribution among the services of the subscription-based CSPs }
\label{tab:csp_user_distributionl}
\end{table}

\noindent \textbf{Profit per engagement time.} 
Based on the 2014-Q2 financial reports, the net income for Google ~\cite{google_revenue2010}, Facebook ~\cite{facebook_revenue2010}, Yahoo ~\cite{yahoo_revenue2010} and AOL ~\cite{aol_revenue2010}, were 3422 M\$, 791 M\$, 272 M\$ and 28.2M\$ respectively. For Microsoft its net income was 432 M\$, obtained by its 2013-Q2 financial report ~\cite{microsoft_revenue2010}, which is the most recent financial report that explicitly denotes the economic growth of the Online Services Division. 

These numbers correspond to a worldwide advertising audience. As the volume of CSPs' profit is proportional to the number of customers to whom the content is shown, we assume that the profit realized in a country is proportional to the percentage of country's Internet users, \ie, we assume uniform profit sharing across countries. 
By combining an Internet Live Stats report ~\cite{itu_world_users} on the worldwide number of Internet users with the U.S. Internet users in 2012 (around 254,295,536), we estimate the quarterly profits of CSPs in the US from the global profit with a ratio of 0.086 (Table~\ref{tab:content_profit_usa_spain}).

Turning to the subscription-based CSPs, we consider the following providers: Netflix, Amazon Prime and Hulu Plus. Netflix charges its existing clients for \$7.99 per month, while it has recently announced to raise its monthly fees for new streaming subscribers to \$8.99~\cite{variety}. Amazon Prime members pay \$99 per year, or \$8.25  per month, while Hulu Plus costs \$7.99 per month. Furthermore as Hulu Plus earns additional profits by showing advertisements along with its content, we assume an average CPM of \$27.61~\cite{reelseo}, while Netflix and Amazon do not show ads.

\begin{table}[tb]
	\small
\centering
\begin{tabular}{|c||c|}
\hline
Content provider & Profit, USA (M\$)  \\
\hline
\hline
Google & 294.29 \\
\hline
AOL & 2.42 \\
\hline
Microsoft & 37.15 \\
\hline
Yahoo & 23.39 \\
\hline
Facebook & 68.02 \\
\hline
\end{tabular}
\caption{Estimated quarterly profits of CSPs} \label{tab:content_profit_usa_spain}
\end{table}

We use the aformentioned computed profits of the ad-powered CSPs (Table~\ref{tab:content_profit_usa_spain}) to estimate the value of the $\rho(s)a(s)$ product, which is the profit per engagement time for service $s$. 
First, based on the report of IAB \cite{iab_ad_revenue_distribution}, search accounted for 43\%, while digital video for 7\% of the advertisement spendings. We assume 80--20\% distribution of revenues between OSNs and gaming from the 22\% of displayed and rich media ads. Second, we assume that all the CSPs derive the same profit per engagement time given a specific service. Hence, we summarize the profits of the considered CSPs and then share the aggregate profit among the services proportional to the percentages reported in the IAB study. Finally, we divide this profit per service value with the aggregate time the users spend in the service category and the total number of subscribers the A-ISPs have. We present the derived $\rho(s)a(s)$ values in Table~\ref{tab:profit_per_engagement_time}.
	
	%
	
%

\begin{table}[tb]
		\small
	\centering
	\begin{tabular}{|c||c|}
	\hline
	Service & USA (\$/min) \\
	\hline
	\hline
	Video & 0.00092 \\
	\hline
	Online social networks & 0.00181 \\
	\hline
	Search & 0.01002 \\ 
	\hline
	Gaming & 0.00092 \\ 
	\hline
	\end{tabular}
	\caption{Estimated profit per engagement time} \label{tab:profit_per_engagement_time}
\end{table}

\noindent \textbf{Estimation of stickiness for ad-powered CSPs.} We estimate the stickiness of the ad-powered CSPs using historical data on their market shares. For user-centric video, we use data of Nielsen~\cite{nielsen_video2012} on the top U.S. online video sites (2009--13), while for search, we use the reports of ComScore~\cite{comscore_search2012} on the U.S. search engine rankings (2008--13).

For each CSP, we compute the ratio of the maximum and the minimum of their market shares. 
This ratio is a worst-case estimation of the stickiness of the CSP. We use $\underline{\theta}=0.36$, AOL's ratio with respect to search, as the lower bound of the CSPs' loyalty, while $\overline{\theta}=0.80$ as upper bound (the ratio of Microsoft in case of video).

\subsection{A-ISPs}
 We use publicly available reports on the US Internet access market to capture the number of end users of each A-ISP (Table~\ref{tab:subscribers_us}).

	\begin{table}[tb]
		\small
	\centering
	\begin{tabular}{|c||c|}
	\hline
	Access ISP & Number of subscribers\\
	\hline
	\hline
	Comcast & 19,025,000 \\
	\hline
	Time Warner & 11,306,000 \\
	\hline
	Cox & 4,590,000 \\
	\hline
	Charter & 3,917,000 \\
	\hline
	Cablevision & 3,060,000 \\
	\hline
	others & 3,872,800 \\
	\hline
	\end{tabular}
	\caption{Access ISPs and their market shares in the US, Q3 2012 \protect\cite{leichtman2012}} \label{tab:subscribers_us}
	\end{table}
	
%
\subsubsection {Modelling assumptions}
\noindent \textbf{A-ISP stickiness.} We distribute the different types of customers across the A-ISPs uniformly, while there is no correlation between the usage patterns of the different services. Hence, all the A-ISPs have initially the same type distribution---but with different size of customer base. In any case, the stickiness of the AISPs is identical: $\beta(i)=\beta, \, \forall i \in ISP$.

\subsubsection {Parameterization assumptions}
\vspace{4pt}
\noindent \textbf{A-ISPs' number of subscribers and annual profits. } 
We derive the profit per customer for A-ISPs as the difference between the price of a subscription and the cost of provisioning it. A report of ITU gives the average price of broadband access~\cite{itu_revenue}: \$20.0 for the US (in 2010).
The average provisioning cost of a broadband connection was 46.45\% of its price in 2010~\cite{cost_of_access_cmt}.
From the above data we compute the annual profits of the A-ISPs, assuming that end-users pay their monthly Internet access fee, in order to use any of the vast array of services offered in the current Internet. As the services offered by the considered CSPs are obviously only a subset of the online services, we deduct that there is a portion of the total annual A-ISPs' profits generated due to the presence of these CSPs and their menu of services. The more significant is the service, based on the engagement time end-users spent on that service (see section 5.1.2 for details), the higher is the portion of the perceived A-ISPs' profits that is attributable to that service. For example, as osn services are more significant than search (based on Table~\ref{tab:services_time}), it appears that a larger portion of the total annual A-ISPs' profits is generated due to the presence of osn services in comparison with the profits due to search services.

\noindent \textbf{Estimation of A-ISPs' stickiness.} We estimate the stickiness of A-ISPs using Leichtman Research Group's historical data on the US broadband market (2009--13)~\cite{leichtman2012}.
For each A-ISP, we compute the ratio of the maximum and the minimum of their market shares based on the historical data. This ratio is a worst-case estimation of the stickiness of the A-ISP. We use $\underline{\beta}=0.77$, Cablevision's ratio, as the lower bound of the A-ISPs' loyalty, while $\overline{\beta}=0.95$ as upper bound (the ratio of Time Warner).



\subsection{General}
\subsubsection {Modelling assumptions}
\vspace{4pt}
\noindent \textbf{Market conditions.} Initially, we assume that the market lacks premium peering agreements: $q(i,x)=0, \, \forall i \in ISP, \forall x \in CSP$, and that that both AISPs and CSPs face the same transit cost: $t(i) = t(x) = t, \, \forall i \in ISP, \forall x \in CSP$.

\subsubsection {Parameterization assumptions}
\noindent \textbf{Peering and transit costs.} Based on fiercewireless.com ~\cite{trans_price}, we use 1K \$ per Gbps/month as transit cost.
We estimate the cost of peering based on the publicly advertised prices of an Internet eXchange Point (IXP) \cite{espanix}. Specifically, the yearly fees are: 
yearly fee of \$2700, and fees \$4000 (FE), \$4500 (GB) and \$14000 (10G) per port per year. 
We assume that in case of a premium peering agreement, at least one additional port should be allocated at each parties (\ie, direct peering).

\section{QUANTITATIVE EVALUATION} \label{sec:evaluation}

In this section we feed data from Sect.~\ref{sec:parameterization} into our model, in order to investigate qualitatively the volume and the direction of payments in the U.S. Internet market. Initially, we concentrate on the Comcast-Google pair (largest A-ISP and ad-powered CSP, respectively), assuming that they are the first to introduce premium peering into the U.S. market. Our experiments reveal that three parameters have the most pronounced impact on the prices of premium peering: Namely, (i) the stickiness of the A-ISPs ($\beta$), (ii) the stickiness of the CSPs ($\theta$), and (iii) the magnitude of the increase of the engagement times ($a \rho$), affect the characteristics of the ecosystem the most. We demonstrate this by implementing several market scenarios, starting from simple ones and gradually introducing more parameters. After showing in what way the various model parameters affect the derived premium peering payments, we conclude with the equivalent price per Gbps per month for each service separately and compare it with the current transit and paid peering prices. Moreover, in order to get a more integrated view of how our model and the predicted payments correlate with actual market strategies, we extend our study with the Comcast-Netflix pair, as they have recently agreed on a paid peering deal. 
 
\subsection{The effect of engagement time increase \& customers churn}

We initially assume that only one service is included in the bilateral premium peering agreement, either search or user-centric video. This permits us to derive interesting conclusions around individual services. Later on we generalize and consider all services in parallel.

\textbf{Zero churn on both sides of the market.} If customers
are fully loyal to both their ISP and CSP, then a premium peering between Comcast and Google is not going to attract any customers from their respective competitors. The only benefit of the premium peering will be an increase of engagement time for their existing customers. While flat monthly payments do not permit Comcast to monetize this increase of engagement time, Google is in position to earn additional profits, and this induces the Nash bargaining solution to transfer some of the benefit over to Comcast. The exact amount depends on the magnitude of increase of the engagement time. We examine two cases, a conservative one, in which the engagement time for video increases by 7.48\% and for search by 0.2\% (as per the reports of ~\cite{akamaiIMC2012,google_latency,microsoft_search_latency}), and an optimistic one, in which the engagement time doubles as a result of the establishment of premium peering. Of course the interesting case here is the second, since there is little sense in discussing payments if the premium peering has a small impact on end user QoE and engagement time. 

Under the conservative estimation, the monthly payment from Google to Comcast for video
is \$11814, while for search is  \$2673, reflecting the fact that the assumed increase of engagement time is much higher for user-centric video. In the optimistic case,
the monthly payment for video is \$0.18M and for search \$1.17M. Notice that in this case, search yields higher payments than user-centric video.  Having equalized the increase of engagement time for both services (doubled), search is carrying a much higher \$ revenue per engagement time increase than video (\$0.01 per extra minute for search versus \$0.00092 per extra minute for user-centric video according to Table~\ref{tab:profit_per_engagement_time}).

\textbf{Customer churn on one side of the market.} Next we demonstrate the effect of customer churn on the direction and the volume of payments.

We start by considering non-zero customer churn on the content side of the market, while keeping the access side churn free.
This means that by establishing a premium peering, Google can attract a part of its competitors' customers, whereas Comcast cannot. 
Having two sources of extra benefit, induces the Nash solution to transfer a larger amount of money from Google to Comcast (for both search and video), compared to the previous case in which the increased engagement time was the only extra benefit. The exact amount depends on the joint effect of the engagement time increase, and loyalty of end-users towards their CSP parameters. For example, when the CSPs' loyalty is equal to 0.4, the monthly payment for user-centric video services from Google to Comcast is \$0.32M, while for CSP's loyalty=0.8, it is down to \$0.23M. In the second case, Google can attract fewer customers from its competitors and thus its benefit, and consequent payment to Comcast is smaller (both amounts are derived under the optimistic case of engagement times increase). 

\begin{figure}[!htbp]
\vspace{-19pt}
\centering
     { \includegraphics[width=75mm]{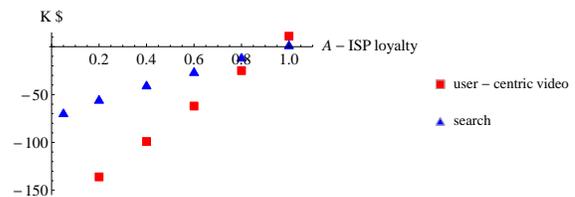}}
  \caption{Monthly payments transferred between Comcast and Google, for zero CSP churn, while customer churn may occur on the access side. Comcast has to pay Google, unless the A-ISP's loyalty is very high.}
	\label{figure:2dfk}
	\vspace{-19pt}
\end{figure}

Next we invert the picture and consider non-zero churn on the ISP side, and zero churn on the content market side. Now Comcast is able to gain a portion of its competitors' customers, while Google benefits only through the increase of engagement time of its existing customers.
The direction of payment now is not strictly from CSP to ISP but it can be the other way around if the incoming churn of the ISP is more beneficial than the increase of engagement time for the CSP. For example, under the conservative case for the increase of engagement time mentioned before ~\cite{akamaiIMC2012,google_latency,microsoft_search_latency}, Comcast has to pay Google, for all but the highest values of A-ISP loyalty as depicted in Fig.~\ref{figure:2dfk}.
\medskip
\medskip

 \medskip
\medskip

In the optimistic case, however, where the engagement time doubles, Google's profits mask the profits of Comcast from incoming churn and thus the payments are from the former to the latter. For example, the monthly payment for user-centric video from Google to Comcast is \$0.10M for A-ISP loyalty equal to 0.4. For ISP loyalty equal to 0.8 the payment increases to \$0.15M, since in this case Comcast attracts fewer customers from competitors and thus the Nash solution assigns it a bigger chunk of Google's benefits from increased engagement.

\textbf{Customer churn on both sides of the market.} We now assume that both Google and Comcast are in position to attract new customers as a result of their premium peering agreement. We still consider a single service, either search or user-centric video.

\begin{table}[H]
\small
\centering
  \begin{tabular}{ | c || c | c|}
    \hline
    CSP's loyalty & M\$(search) & M\$(user-centric video) \\
\hline
\hline 
    0.0 & 1.77  & 0.34  \\ \hline
    0.2 & 1.69  & 0.29  \\ \hline
		0.4 & 1.59  & 0.25  \\ \hline
		0.6 & 1.48  & 0.20  \\ \hline
		0.8 & 1.38  & 0.15  \\ \hline
		1.0 & 1.27  & 0.11  \\ \hline
   \end{tabular}
 \caption{Monthly payments from Google to Comcast for A-ISP's loyalty equal to 0.5. Customers' loyalty towards their CSP impacts the realized payments.}
\label{table:general}
\end{table}

The direction and volume of payments depend on the interplay of all three parameters (engagement time increase and two loyalty parameters). Fixing A-ISP loyalty to 0.5, Table \ref{table:general} shows that for an optimistic end-users' engagement times increase, payments are from Google to Comcast but decrease as the loyalty to CSP increases. The reason is that Google attracts customers of its competitors, alongside a significant increase of its advertising revenues. These benefits outweigh the benefits of Comcast from incoming churn, and thus the Nash bargaining imposes payments from Google to Comcast.  However, as the number of customers attracted by Google
decreases, due to increased CSPs' loyalty, the monthly payments to Comcast decrease as well.

\begin{figure}[!htbp]
  \centering
    {  \includegraphics[width=75mm]{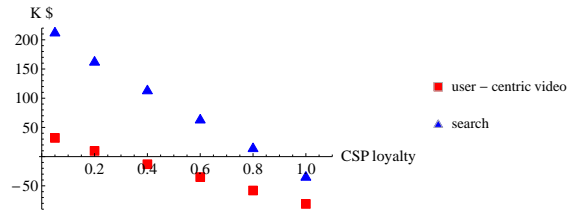}}
		 \caption{Monthly payments for search and user-centric video for low engagement time increase. In most cases Google pays Comcast, although a free-peering agreement could be proved fair for fully CSPs' loyalty.}
\label{figure:1k} 
\end{figure}

Figure \ref{figure:1k} covers the case of pessimistic increase 
of engagement time. Now the extra advertising profits for Google are small and this reduces the volume of payments. At a certain point of CSP loyalty, the benefits from incoming churn in the ISP and CSP counter-balance each other, leading to small payments. A free-peering agreement make sense in this case.

\textbf{The general case: churn in both sides and multiple types of service.} Having examined progressively the effects of engagement time increase and churn on individual services, we next move into the more general and realistic case, in which multiple types service (search, video, osn and gaming) are impacted by the premium peering.  

Qualitatively, all our previous observations remain valid also under this more general case. The volume of payments, however, is now higher in absolute value since it is the result of a summation of multiple side-payments for each one of the considered services. Instead of doing a parameter space exploration with respect to loyalty, we will now attempt to derive payments under loyalty levels extracted from the real US market. 


We compute the monthly payments between Comcast and Google for the four combinations of the extracted loyalty bounds (Sect.~\ref{sec:parameterization}). We use $\underline{\theta}=0.36$, as a lower bound of the CSPs' loyalty, and $\overline{\theta}=0.80$ as an upper bound. Likewise, we assume $\underline{\beta}=0.77$, and $\overline{\beta}=0.95$ as lower and upper A-ISPs' loyalty bounds respectively.

The results of Table \ref{table:generaltable1} indicate monthly payments from Google towards Comcast in the order of \$8.62M--\$9.8M under all combinations of loyalty for the optimistic increase of engagement time. In Set.~\ref{subsec:realbw} we translate these payments into equivalent bandwidth prices. 

An important observation on these payments is that they are actually lower bounds on the amount that the CSP should pay. The reason is that our parameterization in Sect.~\ref{sec:parameterization} has considered only interconnection costs for the A-ISP but no further costs from the increased traffic inside the access and metro networks. Using a CDN co-located with customers (\eg, at DSLAMs) can effectively serve the traffic with little impact on the access (beyond the last mile). In that case, both the A-ISP and the CSP would have close to zero IXP costs, whereas the A-ISP would have to incur the costs of provisioning the CDN infrastructure. We have executed this scenario using a CDN cost of \$4K per Gbps per month.\footnote{\url{http://blog.streamingmedia.com/2012/09/cdn-pricing-stable-survey-data-shows-pricing-down-15-this-year.html}} Table \ref{table:generaltable2} shows that the payments from Google to Comcast are somewhat higher for this superior service ( \$9.07M--\$10.02M per month).

\begin{table}[ht]
\centering
\begin{minipage}[b]{0.45\linewidth}
\begin{tabular}{ r|c|c| }
\multicolumn{1}{r}{}
 &  \multicolumn{1}{c}{$\underline{\theta}$}
 & \multicolumn{1}{c}{$\overline{\theta}$} \\
\cline{2-3}
$\overline{\beta}$ & 10.1 \$ & 9.49 \$ \\
\cline{2-3}
$\underline{\beta}$ & 9.58 \$ & 8.92 \$\\
\cline{2-3}
\end{tabular}
\caption{Million \$ paid by Google to Comcast (per month), without CDN services}
\label{table:generaltable1}
\end{minipage}
\quad
\begin{minipage}[b]{0.45\linewidth}
\begin{tabular}{ r|c|c| }
\multicolumn{1}{r}{}
 &  \multicolumn{1}{c}{$\underline{\theta}$}
 & \multicolumn{1}{c}{$\overline{\theta}$} \\
\cline{2-3}
$\overline{\beta}$ & 10.8 \$ & 10.1 \$ \\
\cline{2-3}
$\underline{\beta}$ & 10.3 \$ & 9.6 \$\\
\cline{2-3}
\end{tabular}
\caption{Million \$ paid by Google to Comcast (per month), with CDN services}
\label{table:generaltable2}
\end{minipage}
\end{table}



\subsection{Impact of A-ISP's market share}
Next we investigate the impact of the size of an A-ISP's customer base on its paid peering relationship with CSPs. Hence, we compare the previous results involving Comcast and Google with a new set of results between Cablevision and Google. Comcast is the biggest player in the US residential broadband marke, while Cablevision is the smallest. In both cases we assume that no previous premium peering agreements exist in the local market.

\begin{figure}
\centering
   \includegraphics[width=75mm]{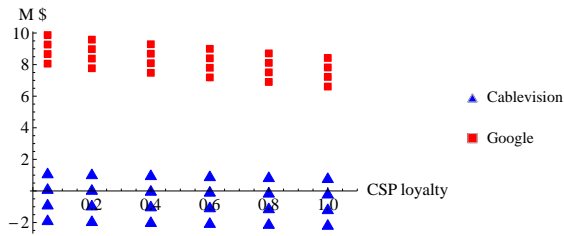}
  \caption{Volume of monthly fees for  Comcast-Google and Cablevision-Google pairs, against multiple values of CSP's loyalty. The large A-ISP always gets the money, while the direction of the money flow between Google and Cablevision strongly depends on the A-ISP's loyalty.}
	\label{figure:2k}

\end{figure}

As Comcast already holds most of the market, a smaller A-ISP has much more to gain from a premium peering relationship with an important CSP. Therefore, whereas before payments went consistently from Google to Comcast, in the case of Cablevision, the ISP can pay the CSP (for low ISP loyalty that permits for maximum incoming churn) and even if it gets paid (under high ISP loyalty that permits less incoming churn), the volume of payment is significantly smaller than the corresponding one for Comcast under similar market conditions. Figure~\ref{figure:2k} presents concrete \$ values for the above observations for our US case study.

Notice that our results  verify real market trends that have large A-ISPs receiving payments from CSPs for direct peering whereas small A-ISPs offering it for free.

\subsection{Timing matters}
\begin{figure}[!htbp]
  \centering
     { \includegraphics[width=75mm]{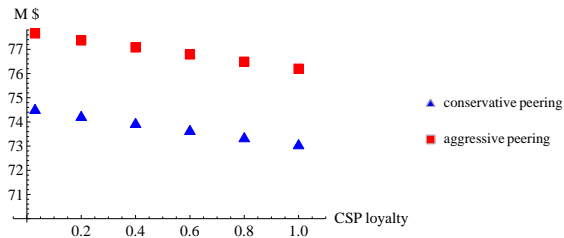}}
  \caption{Comcast's monthly profits for an aggressive vs a conservative A-ISP's peering policy. The post peering profits are decreasing as Comcast becomes more reluctant to establish the premium peering deal.}
	\label{figure:3k}
\end{figure}

We implement a scenario where Comcast is the last A-ISP to agree with Google, and compare it with the  case in which Comcast and Google are the first to establish a premium peering agreement. 
If Comcast acts in the first place (aggressive peering), then it is able to attract a significant portion of its competitors' customers, while late agreements (conservative peering) is more of a defensive measure for retaining its customers, as its competitors have already exploited the privileges of premium peering deals. If this is the case, Comcast's customer base has been decreased due to the migration of some of its end-users to another, more aggressive, access provider. Hence the post peering profits of Comcast will be smaller in comparison with the case in which it acts proactively, as depicted in Fig.~\ref{figure:3k}.  

\begin{figure}[!htbp]
  \centering
     { \includegraphics[width=75mm]{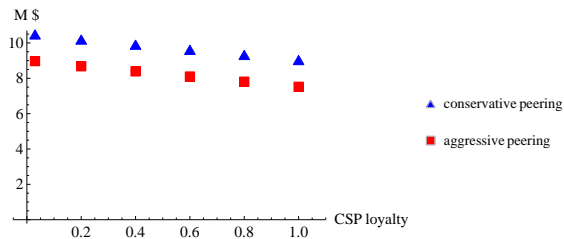}}
  \caption{Monthly charges from Comcast to Google for an aggressive vs a conservative A-ISP's peering policy.}
	\label{figure:3k1}
\end{figure}

Nevertheless, Google is able to increase its customers' population by attracting a portion of Comcast's end-users, which used to interact with another CSP, before the premium peering deal. Hence the direction of payments is still from the CSP to the A-ISP. Interestingly enough, the amount of payments from Google to Comcast for the conservative peering policy is higher  in comparison with the aggressive one. This is justified by the fact that as Comcast delays the premium peering deal and hence it is not in position to attract new customers, Google has to pay higher premium peering fees in order to offer the appropriate incentives to Comcast to accept the deal (see Fig.~\ref{figure:3k1}). 

\subsection{Comparison with real bandwidth prices}\label{subsec:realbw}
Up to now, we have used as metric the monthly payment (in \$) from one party to the other. Next we will compute an equivalent price per Gbps per month for each one of the considered services. The transformation involves just a division of the actual payment reported before, by the extra volume of data transmitted because of the premium peering. The purpose of this study is to see how our payments correlate with actual bandwidth prices in the market.

Based on our most realistic scenario, and assuming that the pair Comcast-Google is the first to establish a premium peering deal, we demonstrate in  Table \ref{table:peerprice2} the equivalent price per Gbps per month for each service (for a significant engagement times' increase).

\begin{table}[H]
\small
\centering

  \begin{tabular}{ | c || c | c| c| c|}
    \hline
    CSP's loyalty & user-centric video & search & osn & gaming \\
\hline
\hline 
		0.0 & 0.96 & 1355 & 69 & 0.22 \\ \hline
 		0.2 & 0.87 & 1276 & 62 & 0.19 \\ \hline
    0.4 & 0.77 & 1196 & 54 & 0.17 \\ \hline
		0.6 & 0.67 & 1116 & 47 & 0.14 \\ \hline
		0.8 & 0.56 & 1035  & 38 & 0.11 \\ \hline
		1.0 & 0.45 & 954  & 30 & 0.08 \\ \hline	
  \end{tabular}
   \caption{K \$ per Gbps/month transferred between Comcast and Google for A-ISP's loyalty equal to 0.95.}
  \label{table:peerprice2}
	
\end{table}

Our first important observation is that \emph{the premium peering prices that we compute for video are reasonably close to real bandwidth price in the market}. Indeed, assuming that paid peering prices are in many cases set to less that the half the price of the corresponding transit prices \footnote{See for example: {http://drpeering.net}}, and based on current market reports, which claim that transit prices in 2013 have dropped below 1 K \$ per Gbps/month ~\cite{trans_price} , we observe that our results are in the same order of magnitude with both the current transit and paid peering prices.

The fact that video premium prices match the above real bandwidth prices is very important for the validity of our model, as video dominates traffic on the Internet. Therefore, if the model and its parameterization are sufficiently representative of reality, then the predicted prices for video should correlate with the real prices for amorphous (\emph{i.e.,} independent of service) bandwidth on the market, which seems to be the case according to our results.

The next column of Table \ref{table:peerprice2}, which corresponds to search, leads us to a second important observation. \emph{The predicted fair prices for search are several orders of magnitude higher than the corresponding prices for video and the real transit and paid peering prices of the month}. This might appear surprising but it is actually consistent with everything else. Video has high volume and low supporting revenue stream, whereas search has low volume and a high supporting revenue stream. A direct consequence of this observation is that \emph{currently CSPs are paying ISPs only for the bandwidth heavy and low profit service (video), but get to enjoy the delivery of their high profit service (search) almost for free}.    

In the following subsection, we like to examine the stability of our main conclusion, e.g., that CSPs tend to pay only for the high volume services, by investigating the derived bandwidth prices for commercial-based video services. We focus on the Comcast-Netflix pair, and we again assume that they are the first A-ISP and subscription-based CSP to establish a premium peering deal.
In the following subsection, we like to examine the stability of our main conclusion--that CSPs tend to pay only for the high volume services, by investigating the derived bandwidth prices for commercial-based video services. We focus on the Comcast-Netflix pair, and we again assume that they are the first A-ISP and  CSP pair  establish a premium peering deal.

\subsection{Bandwidth prices for the Comcast-Netflix pair}\label{subsec:realbw2}

The Comcast-Netflix pair is of special interest, as they have recently signed a paid-peering agreement. Although there are several (and in many cases controversial) reports attempting to estimate the exact bandwidth price derived by that deal \footnote {{http://blog.streamingmedia.com/2014/02/heres-comcast-netflix-deal-structured-numbers.html}}, the involved parties have not published any official report of their agreement. As we have already investigated in what way the various model parameters affect the direction and the amount of payments, we proceed to the estimation of the \$ per Gbps per month, transferred between Comcast and Netflix, and present our results in Table \ref{table:peerprice3}.

Our results summarized in Table \ref{table:peerprice3}, are lower than the predicted bandwidth prices for user-centric video (see Table \ref{table:peerprice2}), and consequently lower than the current transit and paid peering prices. This observation is a natural outcome of our model, as Netflix is not an ad-powered CSP and hence can not be benefited by the increase of the end-users' engagement times. Both Comcast and Netflix have only one source of additional income i.e., the customers' churn on each side of the market, after the premium peering deal.

\begin{table}[tb]
\small
\centering
  \begin{tabular}{ | c || c | }
    \hline
    CSP's loyalty & commercial based video  \\
\hline
\hline 
		0.0 & 0.18 \\ \hline
 		0.2 & 0.15  \\ \hline
    0.4 & 0.12  \\ \hline
		0.6 & 0.08 \\ \hline
		0.8 & 0.05 \\ \hline
		1.0 & 0.01  \\ \hline	
  \end{tabular}
   \caption{K \$ per Gbps/month paid by Netflix to Comcast for A-ISP's loyalty equal to 0.95.}
  \label{table:peerprice3}
\end{table}
The bandwidth prices, presented in Table \ref{table:peerprice3}, do not take into account that a paid peering agreement between an A-ISP and a CSP, includes far more complex investment decisions than a simple acquisition of the dedicated ports and links. As highlighted in ~\cite{Dovrolis2014}, direct interconnection agreements require additional investments, capacity upgrades and more sophisticated routing policies especially by the A-ISP, which owns the most capital-intensive part of the network. Indeed, the direct interconnection between two providers is ultimately an agreement to share their entire infrastructures, and not simply a link and some router ports.
Motivated by this observation and similarly to our general case analysis, we present in ( Table \ref{table:peerprice4}), our bandwidth prices for the Comcast-Netflix pair, when the A-ISP uses CDN services in the last-mile, in order to deliver the content effectively to the end-users.   

\begin{table}[tb]
\small
\centering
  \begin{tabular}{ | c || c | }
    \hline
    CSP's loyalty & commercial based video  \\
\hline
\hline 
		0.0 & 0.31 \\ \hline
 		0.2 & 0.29 \\ \hline
    0.4 & 0.27  \\ \hline
		0.6 & 0.24 \\ \hline
		0.8 & 0.21 \\ \hline
		1.0 & 0.19 \\ \hline	
  \end{tabular}
   \caption{K \$ per Gbps/month paid by Netflix to Comcast for A-ISP's loyalty equal to 0.95, with CDN services}
  \label{table:peerprice4}
\end{table}

In this case the predicted bandwidth prices are higher, that without the use of CDNs, and mush closer to the empirical rule, that paid peering prices are half the price of the transit fees.

Another interesting observation is that the payments (as \$ per Gbps/month) transferred from Netflix to Comcast are less sensitive to the CSP stickiness, in comparison with the user-centric video case. This is justified by the fact that Netflix is by far the dominant player in the subscription-based video market, and hence the incoming population after the premium peering deal will be any case relatively small. Consequently, as the CSP customer churn will be negligible, it won't affect significantly the derived payments.

The aforementioned conclusions substantiate our claim that per service peering might
provide a transcendent framework on how a fair peering price could be determined. In fact, there is an ongoing work on enabling "application-specific peering", by deploying Software Defined Networking (SDN) at the various Internet eXchange Points (IXPs) ~\cite{SDX2014}. This architecture is aiming to provide each ISP (either A-ISP or CSP) the flexibility to express direct expression of more sophisticated interconnection strategies, such as the per service peering. Hence, our work provides another perspective on the emergence of application-specific peering, by presenting a quantitative analysis of the endogenous asymmetries of the various types of services.

Finally, based on our analysis, we have developed \emph{PeeringCalc}, a web-site that permits users
to evaluate our model, on their own case-study. (URL: http://peeringcalc.nes.aueb.gr)

\include*{related_work}
\section{Conclusions and Future Work}


The main disagreement between CSPs and A-ISPs is who eventually benefits more from it, with CSPs claiming that improved QoE benefits the A-ISP via positive incoming churn from competitors, and A-ISP complaining about the high costs for delivering it, and the imbalance between their flattening revenues and the higher margins of CSPs. In this work we have attempted to create a model that connects directly all the ingredients of the tussle: the profits from incoming churn in both sides of the market, the profitability of improved QoE, per service characteristics, including loyalty, and the costs of interconnection (peering, CDN, \etc). Unlike all previous work in the area our model derives actual bandwidth prices that can be contrasted with real bandwidth prices on the market and thus serve as a benchmark for actual negotiations and/or regulation.         

Our model cannot obviously capture every detail of reality but we have put special effort in consulting with industry insiders (both in ISP and CSP side, as well as regulators) to make sure that we capture at least the basic dynamics of the tussle. The fact that we output prices that follow intuition qualitatively, and, in some (important) cases match real bandwidth prices from the market, is in our opinion an indication that the model is useful and moving to the right direction. There are of course many improvements that we would like to pursue as well as additional case studies to be conducted. For example, we would like to encode more detailed cost structures to the model and additional data around them. Our definition of loyalty can obviously benefit from additional experimentation in the area and similarly for the the engagement time parameter. In terms of case studies, we are currently developing a web-site that will permit actual stakeholders to populate our model with their own data. We are also considering adapting the model to the wireless broadband market and consider the effects of competition where user prices have more complex structures.

\bibliographystyle{plain}
\bibliography{references}

\end{document}